\documentclass[aps,prl,twocolumn,superscriptaddress]{revtex4}
\usepackage{graphicx}
\usepackage{amssymb}
\usepackage{pifont}

\begin{document}



\title{Possibility of valence-fluctuation mediated superconductivity in Cd-doped CeIrIn$_5$ probed by In-NQR}

\author{M.~Yashima}
\affiliation{Department of Materials Engineering Science, Osaka University, Osaka 560-8531, Japan}
\affiliation{JST, TRiP (Transformative Research-Project on Iron Pnictides), Chiyoda, Tokyo 102-0075, Japan}
\author{N.~Tagami}
\affiliation{Department of Materials Engineering Science, Osaka University, Osaka 560-8531, Japan}
\author{S.~Taniguchi}
\affiliation{Department of Materials Engineering Science, Osaka University, Osaka 560-8531, Japan}
\author{T.~Unemori}
\affiliation{Department of Materials Engineering Science, Osaka University, Osaka 560-8531, Japan}
\author{K.~Uematsu}
\affiliation{Department of Materials Engineering Science, Osaka University, Osaka 560-8531, Japan}
\author{H.~Mukuda}
\affiliation{Department of Materials Engineering Science, Osaka University, Osaka 560-8531, Japan}
\affiliation{JST, TRiP (Transformative Research-Project on Iron Pnictides), Chiyoda, Tokyo 102-0075, Japan}
\author{Y.~Kitaoka}
\affiliation{Department of Materials Engineering Science, Osaka University, Osaka 560-8531, Japan}
\author{Y. \=Ota}
\affiliation{Department of Physics, Graduate School of Science, Osaka University, Osaka 560-0043, Japan}
\author{F.~Honda}
\affiliation{Department of Physics, Graduate School of Science, Osaka University, Osaka 560-0043, Japan}
\author{R.~Settai}
\affiliation{Department of Physics, Graduate School of Science, Osaka University, Osaka 560-0043, Japan}
\author{Y.~\=Onuki}
\affiliation{Department of Physics, Graduate School of Science, Osaka University, Osaka 560-0043, Japan}

\begin{abstract}
We report on a pressure-induced evolution of exotic superconductivity and spin correlations in CeIr(In$_{1-x}$Cd$_{x}$)$_5$ by means of In-Nuclear-Quadrupole-Resonance (NQR) studies. Measurements of an NQR spectrum and nuclear-spin-lattice-relaxation rate $1/T_1$ have revealed that antiferromagnetism induced by the Cd-doping emerges locally around Cd dopants, but superconductivity is suddenly induced at $T_c$ = 0.7 and 0.9 K at 2.34 and 2.75 GPa, respectively. The unique superconducting characteristics with a large fraction of the residual density of state at the Fermi level that increases with $T_c$ differ from those for anisotropic superconductivity mediated by antiferromagnetic correlations. By incorporating the pressure dependence of the NQR frequency pointing to the valence change of Ce, we suggest that unconventional superconductivity in the CeIr(In$_{1-x}$Cd$_{x}$)$_5$ system may be mediated by valence fluctuations.
\end{abstract}
\vspace*{5mm}

\maketitle
\begin{figure}[t]
\centering
\includegraphics[width=7cm]{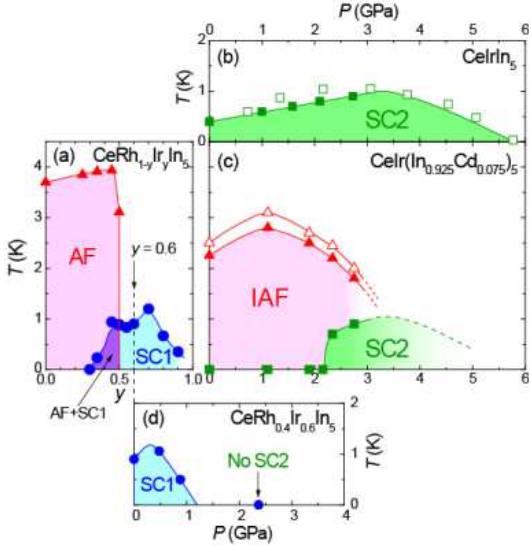}
\caption[]{\footnotesize (Color online) (a) The phase diagram for CeRh$_{1-x}$Ir$_{x}$In$_5$ as a function of Ir concentration. $T_N$ and $T_c$ are referred from Refs.~\cite{Zheng,SKawasaki2}. SC1 is expected to be mediated by magnetic correlations. The length of the horizontal axis between $y$ = 0 and 1 is adjusted to coincide with that between 0 and 3 GPa in the other figures.  (b) The pressure($P$)-temperature($T$) phase diagram for CeIrIn$_5$ \cite{SKawasaki1}. The open squares are referred from the resistivity measurement \cite{Muramatsu}. SC2 is expected to be mediated by valence fluctuations. (c) The $P - T$ phase diagram for CeIr(In$_{0.925}$Cd$_{0.075}$)$_5$ which was determined by the present experiment. The $T_N$(solid ) and  $T_N'$(open triangles) are determined from the peak in the $T$ dependence of $1/T_1$ and the broadening of NQR spectrum due to the onset of the AFM order, respectively. (d) The $P-T$ phase diagram for CeRh$_{0.4}$Ir$_{0.6}$In$_5$ obtained from the $T_1$ measurements.}
\label{phase}
\end{figure}

Cerium (Ce)-based heavy-fermion compounds, CeMIn$_5$ (M = Co, Rh, and Ir), provide us with the opportunity to systematically investigate the interplay between antiferromagnetism (AF) and superconductivity (SC) \cite{Petrovic1,Hegger,Muramatsu,Petrovic2}.  CeIrIn$_5$ and CeCoIn$_5$ show SC at ambient pressure ($P$) below $T_c$ = 0.4 and 2.3 K, respectively \cite{Petrovic1,Hegger,Muramatsu,Petrovic2}. In CeRhIn$_5$, which is an antiferromagnet with $T_N$ = 3.8 K at ambient $P$, SC occurs at $T_c \sim$ 2.2 K under $P$ over 2.1 GPa. However, it has been suggested that the superconducting nature in CeIrIn$_5$ should be distinguished from those in CeCoIn$_5$ and CeRhIn$_5$ from several points of view. The existence of two superconducting phases was reported in CeRh$_{1-y}$Ir$_y$In$_5$ \cite{Nicklas1}. Figs.\ref{phase}(a) and \ref{phase}(b) show the respective phase diagrams of SC (denoted as SC1 and SC2) for CeRh$_{1-y}$Ir$_y$In$_5$ and CeIrIn$_5$ under $P$. SC1 is closely related to antiferromagnetic (AFM) correlations, as argued extensively in CeCoIn$_5$\cite{Yashima1} and CeRhIn$_5$\cite{Yashima2}, whereas it is reported from the previous NQR measurements that the SC2 in CeIrIn$_5$ occurs in a different situation from SC1 as follows \cite{SKawasaki1}. The maximum of $T_c\sim$ 1 K around 3 GPa takes place under the absence of AFM correlations that is suggested by the fact that the $(T_1T)^{-1}$-constant behavior, which is an order of magnitude suppressed by $P$, is observed above $T_c$. Here, $T_1$ is a nuclear-spin-lattice-relaxation time. However, the values of $(T_1T)^{-1}$ are an order of magnitude larger than those in LaIrIn$_5$, indicating that nearly wave-number independent magnetic and/or valence fluctuations are dominant under high $P$. Actually, the large electronic specific heat coefficient $\gamma\sim$ 0.38 J/molK$^2$ at $P$ = 1.56 GPa is observed in CeIrIn$_5$ \cite{Borth}. Moreover, from the resistivity measurements, a non-Fermi liquid behavior that the resistivity approximately follows a $T^{1.5}$ law has been observed in the wide $P$ range of 0 to 3.1 GPa \cite{Muramatsu}. Furthermore, we note that SC2 does not occur in the high $P$ region in CeRh$_{0.4}$Ir$_{0.6}$In$_5$, while SC1 is observed in the low $P$ region, as shown in Fig. \ref{phase}(d). These results suggest that SC2 is discontinuous with SC1 and hence the substitution of Rh for Ir does not correspond to the application of $P$ in CeIrIn$_5$. Therefore, it is anticipated that the superconducting mechanism of SC2 differs from that of SC1.

Meanwhile, it was reported that all the anomalous transport properties observed in CeRh$_{0.2}$Ir$_{0.8}$In$_5$(SC1) and CeIrIn$_5$(SC2) originate from the AFM spin fluctuations irrespective of the superconducting phase to which the system belongs. \cite{Nakajima}. In this context, the reason why the maximum $T_c$ of SC2 in CeIrIn$_5$ is realized far away from an AFM quantum critical point (QCP) is still an underlying issue. The existence of SC1 and SC2 was first reported in CeCu$_2$(Si$_{1-x}$Ge$_x$)$_2$ \cite{Yuan}. It is suggested from extensive experimental and theoretical studies that SC2 emerging in the high $P$ region is mediated by valence fluctuations \cite{Holmes, Fujiwara, Onishi, Watanabe}. In order to develop insight into SC2, we report on the $P$-induced evolution of unique SC characteristics in pure and Cd-doped CeIrIn$_5$.

\begin{figure}[t]
\centering
\includegraphics[width=7cm]{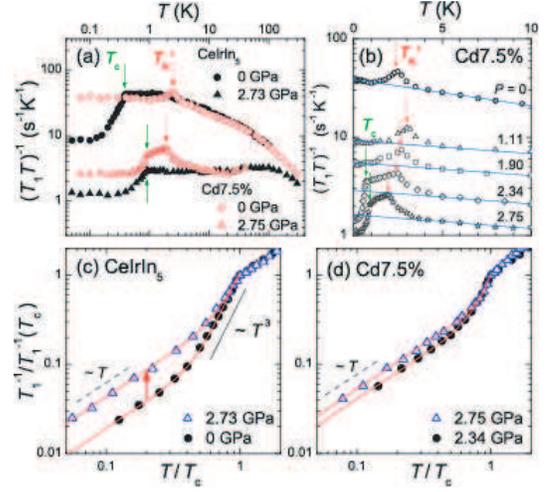}
\caption[]{\footnotesize (Color online) (a) The $T$ dependences of $(T_1T)^{-1}$ for CeIrIn$_5$ and CeIr(In$_{0.925}$Cd$_{0.075}$)$_5$. (b) The $T$ dependences of $(T_1T)^{-1}$ at several pressures in the $P$ range of $0-2.75$ GPa in CeIr(In$_{0.925}$Cd$_{0.075}$)$_5$. To prevent data points from overlapping, the $(T_1T)^{-1}$ results at 2.34 and 2.75 GPa are divided by 1.5 and 2.4, respectively. (c) The $T$ dependence of $1/T_1$ normalized at $T_c$ at $P$=0 and 2.73 GPa in CeIrIn$_5$. (d) The $T$ dependences of $1/T_1$ normalized at $T_c$ at $P$=2.34 and 2.75 GPa in CeIr(In$_{0.925}$Cd$_{0.075}$)$_5$.}
\label{T1}
\end{figure}

Single crystals of CeIr(In$_{1-x}$Cd$_x$)$_5$, CeRh$_{1-y}$Ir$_y$In$_5$, and CeCo(In$_{0.9}$Cd$_{0.1}$)$_5$ grown by the self-flux method were crushed into coarse powder in order to allow the RF pulses to easily penetrate the sample for NQR measurements. Here $T_c$'s for the samples are determined by a sudden decrease in nuclear-spin-lattice-relaxation rate $1/T_1$ below $T_c$. Hydrostatic $P$ was applied by utilizing a NiCrAl-BeCu piston-cylinder cell filled with Daphne 7474 as a $P$-transmitting medium \cite{Murata}. To calibrate $P$ at low temperatures, the shift in $T_c$ of Sn metal at $P$ was monitored by resistivity measurements. CeMIn$_5$ consists of alternating layers of CeIn and MIn$_4$, and there are two sites---In(1) and In(2)---per unit cell. The In(1) and In(2) sites are located in the CeIn and MIn$_4$ layers, respectively. The measurements of an $^{115}$In($I=9/2$)-NQR spectrum and  $1/T_1$ were mainly performed at the transition of 2$\nu_Q$ for the high symmetry In(1) site in CeMIn$_5$. $T_1$ was measured by the conventional saturation-recovery method. When the system is near the AFM QCP, the relations of $(T_1T)^{-1}$ $\propto$ $\chi_Q(T)^n$ ($n =$ 1/2 or 1) for the 2- or 3-dimensional AFM spin-fluctuation models are predicted, respectively \cite{Moriya}. Since the staggered susceptibility follows the Curie-Weiss law as $\chi_Q(T)$  $\propto$ $1/(T+\theta)$, the increase in  $(T_1T)^{-1}$ with decreasing $T$ is observed near the AFM QCP. Here, an NQR frequency ($\nu_Q$) is defined by the NQR Hamiltonian, $\mathcal{H}_Q$ = $(h\nu_Q/6)[3{I_z}^2-I(I+1)+\eta({I_x}^2-{I_y}^2)]$, where $\eta$ is the asymmetric parameter of the electric field gradient [$\eta$ = 0 at the In(1) site].

Figure \ref{T1}(a) shows the $T$ dependence of $(T_1T)^{-1}$ in CeIrIn$_5$ and CeIr(In$_{0.925}$Cd$_{0.075}$)$_5$. The AFM order induced by the Cd-doping into the CeMIn$_5$ system was reported from the specific-heat measurements \cite{Pham}. In fact, the onset of AFM for CeIr(In$_{0.925}$Cd$_{0.075}$)$_5$ is corroborated by the observation of a peak in $(T_1T)^{-1}$ at $T_N \sim$ 2.3 K as shown in Fig. \ref{T1}(a). The $T$ dependence of $(T_1T)^{-1}$ above $T_N$ for the Cd-doped sample almost coincides with that for the pure one. The effect of the Cd-doping would be local in the normal state, as reported in the previous NQR measurements in Cd-doped CeCoIn$_5$ \cite{Urbano}. This result suggests that Cd acts as a local defect that nucleates droplets of local AFM order in a system close to an AFM QCP.

\begin{figure}[htbp]
\centering
\includegraphics[width=7cm]{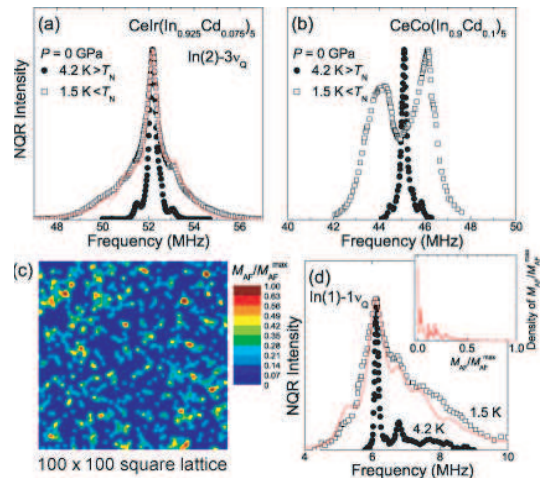}
\caption[]{\footnotesize (Color online) (a) The 3$\nu_Q$ spectra at the In(2) site in CeIr(In$_{0.925}$Cd$_{0.075}$)$_5$. The solid line is the calculated result (see text). (b) The 3$\nu_Q$ spectra at the In(2) site in CeCo(In$_{0.9}$Cd$_{0.1}$)$_5$. (c) Example of the $M_{AF}$ distribution on a $100 \times 100$ square lattice calculated with randomly distributed $7.5 \%$ Cd dopants. (d) The 1$\nu_Q$ spectrum at the In(1) site in CeIr(In$_{0.925}$Cd$_{0.075}$)$_5$. The solid line is the calculated result (see text). The inset shows the density of $M_{AF}/M_{AF}^{max}$ for a $500 \times 500$ square lattice.}
\label{spectrum}
\end{figure}

Figs.\ref{spectrum}(a) and \ref{spectrum}(b) show the NQR spectra for 3$\nu_Q$ at the In(2) site above and below $T_N$ in CeIr(In$_{0.925}$Cd$_{0.075}$)$_5$ and CeCo(In$_{0.9}$Cd$_{0.1}$)$_5$, respectively. A clear splitting into two NQR spectra indicates the occurrence of a homogeneous AFM order with a uniform magnetic moment $M_{AF}$ over the whole sample in Cd-doped CeCoIn$_5$, as presented in Ref. \cite{Urbano}. In Cd-doped CeIrIn$_5$, however, the NQR spectrum below $T_N$ exhibits no clear splitting, but a large broadening at its tail, indicating an inhomogeneous AFM order (IAF) with a large distribution of $M_{AF}$. This difference in the character of the AFM order is probably relevant to the fact that CeCoIn$_5$ is much closer to an AFM QCP than CeIrIn$_5$. In order to explain the spectral shape for IAF, we assume that the magnitude of $M_{AF}$ at each Ce site depends on the density of Cd dopants surrounding it, as shown in Fig. 3c. We adopt a simple square lattice model (500 $\times$ 500) under randomly distributed $7.5\%$ Cd dopants and a spatial distribution of $M_{AF}$ at each Ce site is given by $M_{AF}^{i,j} \propto \sum_{k,l} \exp({-\frac{|\vec{r}_{k,l}-\vec{r}_{i,j}|^2}{\xi^2}})$, where $(i, j)$ and $(k, l)$ indicate positions of Ce and Cd ions, respectively, and $\xi$ is a length that the influence of Cd dopants reaches. Incorporating some local effect of the Cd-doping, a Gaussian function is tentatively assumed to calculate $M_{AF}^{i,j}$. The distribution of $M_{AF}$ obtained from this model with $\xi = 1.4 a$ ($a$ is the lattice parameter of a-axis) is shown in the inset of Fig. 3d. The reproducibility for the distribution of $M_{AF}$ was confirmed due to the sufficiently large size of a $500 \times 500$ square lattice. The simulated spectral shape for 3 $\nu_Q$ at In(2) with $M_{AF}^{max} =$ 8.7 kOe and a linewidth broadening factor $\eta =$ 250 kHz is shown by the solid line in Fig. 3a. The simulation for 1 $\nu_Q$ at In(1) with $M_{AF}^{max} =$ 2.7 kOe and $\eta =$ 300 kHz is also shown by the solid line in Fig. 3d. These simulations are in good agreement with their respective experimental results, indicating that $M_{AF}$ is randomly distributed as illustrated in Fig. 3c.

Next, we discuss superconducting characteristics for CeIr(In$_{0.925}$Cd$_{0.075}$)$_5$. The $T$ dependences of $1/T_1$ normalized at $T_c$ are shown for the pure and Cd-doped samples in Figs. \ref{T1}(c) and \ref{T1}(d), respectively. In general, the $T$ dependence of $1/T_1$ below $T_c$ allows us to estimate a superconducting gap (2$\Delta_0 / k_B T_c$) and a residual density of states (RDOS) at the Fermi level due to some impurity effect by assuming a certain pairing symmetry. Here, we tentatively assume a $d_{x^2-y^2}$-wave model which is indicated by the thermal-conductivity, magnetic-penetration-depth, and specific-heat measurements \cite{Kasahara,Vandervelde,Kittaka}. The solid lines in Figs. \ref{T1}(c) are the calculated results for CeIrIn$_5$ with parameters of (2$\Delta_0/k_B T_c$, RDOS) = (5.2, 0.44) at $P$ = 0 and (5.2, 0.66) at $P$ = 2.73 GPa. Since RDOS in CeIrIn$_5$ are unexpectedly much larger than RDOS = 0.08 at $P$ = 0 in CeCoIn$_5$ \cite{Yashima1} and 0.14 at $P$ = 2.35 GPa in CeRhIn$_5$ \cite{Yashima3} in spite of its high purity, it is possible that a superconducting gap is not formed or is small enough to be easily broken even by weak impurity scattering in one or a few bands of the Fermi surface. Here, we highlight the fact that RDOS in CeIrIn$_5$ increases from 0.44 to 0.66 despite the enhancement of $T_c$ from 0.4 to 0.9 K. Such a behavior contrasts with those in CeCoIn$_5$ and CeRhIn$_5$ in which RDOS remains almost unchanged and is reduced as $P$ increases, respectively. The unconventional enhancement of RDOS by $P$ cannot be understood in terms of a simple impurity effect. Analogous SC characteristics are also observed in the Cd-doped sample with the parameters of (2$\Delta_0/k_B T_c$, RDOS) = (6.4, 0.64) and (7.7, 0.73) at $P$ = 2.34 and 2.75 GPa, respectively, indicating that it is intrinsic in the CeIr(In$_{1-x}$Cd$_{x}$)$_5$ system.

\begin{figure}[htbp]
\centering
\includegraphics[width=6cm]{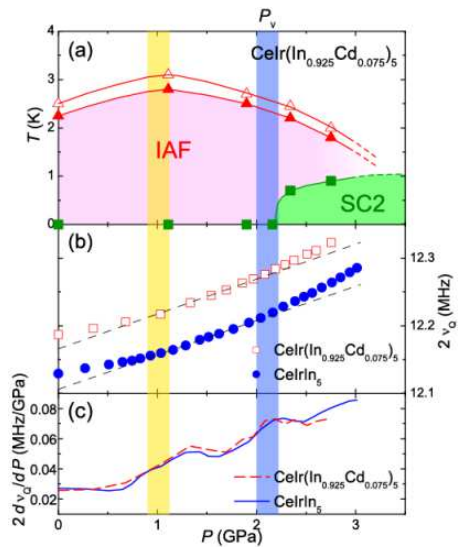}
\caption[]{\footnotesize (Color online) (a) The $P-T$ phase diagram for CeIr(In$_{0.925}$Cd$_{0.075}$)$_5$. (b) The $P$ dependences of 2$\nu_Q$ for the In(1) site in CeIrIn$_5$ and CeIr(In$_{0.925}$Cd$_{0.075}$)$_5$. The dashed lines are guides for the eyes. (c) The $P$ dependences of 2$d\nu_Q/dP$ obtained from Fig.~\ref{nuQ}(b). The values of 2$d\nu_Q/dP$ are smoothed by a simple moving average with the nearest neighbors.}
\label{nuQ}
\end{figure}

Since $\nu_Q$ depends directly on an electric field gradient at observed nuclei, probing a charge distribution emerging from surrounding ions and electrons, its $P$ dependence is relevant with some changes in both the valence of Ce and a lattice volume. A structural phase transition is simply expected as one possible cause for an anomaly in the $P$ dependence of $\nu_Q$, but it is suggested from the x-ray diffraction measurements that no $P$-induced structural phase transition occurs up to 15 GPa in CeIrIn$_5$ \cite{Kumar}. We give an example of an anomaly in the $P$ dependence of $\nu_Q$ in CeCu$_2$Si$_2$ relevant with the valence crossover. The NQR measurements under $P$ in CeCu$_2$Si$_2$ reported that a change in the slope of the $\nu_Q$ versus $P$ curve, $d\nu_Q/dP$ is observed around $P \sim 4.2$ GPa, just below where $T_c$ reaches the maximum \cite{Fujiwara,Holmes}. It is suggested from the recent resistivity measurements that CeCu$_2$Si$_2$ lies in proximity to a valence transition: the critical end point between the first-order valence transition and valence crossover could be located at $4.5\pm0.2$ GPa and a slightly negative $T$ \cite{Seyfarth}. Figure \ref{nuQ}(b) indicates the $P$ dependence of 2$\nu_Q$ for the pure and Cd-doped samples. The $P$ dependences of $\nu_Q$ closely resemble each other with a nearly 60 kHz gap between both samples. As shown in Fig. \ref{nuQ}(b), $\nu_Q$ monotonously increases up to about 1 GPa, and $d \nu_Q/d P$ becomes steep above 1 GPa (see also Fig. \ref{nuQ}(c)). It is noteworthy that $T_N$ starts to decrease beyond 1 GPa, as confirmed in Fig.~\ref{nuQ}(a). This may be caused by the increase in the hybridization between Ce-$4f$ and conduction electrons which triggers a change of the Ce valence above 1 GPa. As $P$ increases further, $d\nu_Q/d P$ also increases around $P_{v} \sim$ 2.1 GPa for both the samples, as shown in Fig.~\ref{nuQ}(b) and \ref{nuQ}(c). Since SC2 suddenly emerges around $P_{v}$ in the Cd-doped sample and the maximum of $T_c$ is realized beyond $P_{v}$, it is likely that the anomaly of $\nu_Q$ at $P_{v}$ is closely related with the Ce valence crossover inducing valence fluctuations as well as that for CeCu$_2$Si$_2$ under $P$ \cite{Fujiwara}. It should be noted that the $P$ variation of $T_c$ around $P_{v}$ in CeIrIn$_5$ is softer than that in CeCu$_2$Si$_2$, indicating that the valence crossover in CeIrIn$_5$ is not rapid, as compared to that in CeCu$_2$Si$_2$. Moreover, it is also expected that the unexpected enhancement of RDOS discussed before may arise from valence fluctuations. This is because the theory points out that valence fluctuations enhance impurity scattering as observed in CeCu$_2$Si$_2$ \cite{Holmes}. In order to clarify the relation between the unexpected enhancement of RDOS and valence fluctuations, further theoretical work on the formation of a superconducting energy gap by valence fluctuations is clearly needed.

Next, we focus on an evolution of low-lying excitations which suddenly emerge above $P_{v}$ in the Cd-doped sample. Fig.~\ref{T1}(b) indicates that the $T$ dependences of $(T_1T)^{-1}$ below $T_N$ are smoothly extrapolated from those well above $T_N$ in $P$ = 0 - 1.90 GPa below $P_{v}$, where SC is not induced. This result suggests that AFM moments around Cd dopants become static below $T_N$ due to the formation of static IAF order because the enhancement in $(T_1T)^{-1}$ due to the AFM order appears only in the vicinity of $T_N$. However, once $P$ exceeds $P_{v}$, the $T$ dependences of $(T_1T)^{-1}$ below $T_N$ are not extrapolated from those well above $T_N$, indicating that the low-lying excitations remaining even well below $T_N$ may be responsible for the onset of SC2. A likely explanation for this unrecovered enhancement of $(T_1T)^{-1}$ is that magnetic moments are still fluctuating even below $T_N$ through the development of valence fluctuations above $P_{v}$. Here, we emphasize that the stronger coupling effect in the Cd doped sample than in the pure one causes a comparable $T_c$ with that in the pure one through the overcoming of a possible reduction of $T_c$ due to the impurity effect and the induced IAF order. In this context, it is expected that the formation of strong coupling SC in the Cd doped sample is closely related to the low-lying excitations reinforced by valence fluctuations. Since the Cd-doping causes some change in the low-energy spin dynamics above $P_{v}$ after all, further theoretical studies on a possible cooperative coupling between valence and spin fluctuations may be necessary to clarify the complicated mechanism of SC2 for the CeIr(In$_{1-x}$Cd$_{x}$)$_5$ system.

In conclusion, the present In-NQR measurements have revealed that SC2 in pure and Cd-doped CeIrIn$_5$ differs from SC1 that is closely related to the AFM correlation in many respects: the $P$ dependence of $\nu_Q$ pointing to the valence change of Ce, the unexpected increase in RDOS by the application of $P$, and the unrecovered enhancement of $(T_1T)^{-1}$ even below $T_N$ above $P_{v}$. These results lead us to consider that unconventional SC2 is likely mediated by valence fluctuations.

We thank S. Watanabe and K. Miyake for enlightening discussions and theoretical comments. This work was supported by Grants-in-Aid for Specially Promoted Research (No. 20001004), for Young Scientists (B) (No. 20740195), and for Scientific Research (C) (No. 24540372) from the Ministry of Education, Culture, Sports, Science and Technology (MEXT) of Japan. It was partially supported by the Global COE Program (No. G10) from MEXT.

\end{document}